\documentclass[12pt,preprint]{aastex}
\newcommand{\be}{\begin{equation}}
\newcommand{\ee}{\end{equation}}
\newcommand{\bbb}{\begin{eqnarray}}
\newcommand{\eee}{\end{eqnarray}}
\newcommand{\ph}{\phantom{0}}

\begin{document}
\shorttitle{Active region seismology}
\title{Ring diagram analysis of the characteristics of solar oscillation modes
in active regions}
\author{S. P. Rajaguru}
\affil{Indian Institute of Astrophysics, Bangalore-560 034, India }
\email{rajguru@iiap.ernet.in}
\author{Sarbani Basu}
\affil{Astronomy Department, Yale University, P. O. Box 208101,
New Haven CT 06520-8101, U.S.A.}
\email{basu@astro.yale.edu}
\and
\author{H. M. Antia}
\affil{Tata Institute of Fundamental Research,
Homi Bhabha Road, Mumbai 400005, India}
\email{antia@tifr.res.in}

\begin{abstract} 
The presence of intense magnetic fields in and around
sunspots is expected to modify the solar structure and oscillation
frequencies.  Applying the ring diagram technique to data from the
Michelson Doppler Imager (MDI) on board the Solar and Heliospheric
Observatory (SOHO), we analyze the characteristics of high-degree f and
p modes near active regions and compare them with the characteristics of
the modes in quiet regions. As expected from earlier results, the f- and
p-mode frequencies of high degree modes are found to be significantly
larger in magnetically active regions. In addition, we find that the power
in both f and p modes is lower in active regions, while the widths of the
peaks are larger, indicating smaller lifetimes. We also find that the
oscillation modes are more asymmetric in active regions than those in
quiet regions, indicating that modes in active regions are excited closer
to the surface. While the increase in mode frequency is monotonic in
frequency, all other characteristics show more complex frequency
dependences. 
\end{abstract} 

\section{Introduction}
\label{sec:intro}

The solar activity cycle is marked by the appearance of sunspots, which
are regions of intense magnetic field. The presence of these magnetic
fields in and around sunspots in and below the photosphere modifies the
equilibrium profile of sound speed and density, through the Lorentz force.
This change in the properties of localized regions in the Sun is expected
to modify the characteristics of solar f  and p modes. Sunspots are known
to scatter and absorb p modes, as found by the early observations
\citep{braunetal87,braunetal88,braunetal90,braun95}. 
Despite a number of subsequent studies in this
field (see e.g., \citet{hollweg88, rosenthal90, bogdanetal93,
keppensetal94,bogdanetal96}),
the exact mechanism of the phenomenon remains unknown. It
is also well known that amplitudes of solar oscillations decrease
significantly in active regions \citep{tarbelletal88,hindman-brown98}. It
is not clear if the decrease in power is due to absorption of acoustic
modes by sunspots or due to weaker excitation in active regions. High
degree modes, i.e., those with very short horizontal wavelengths, are
expected to be affected the most by sunspots. The lifetime of these modes
is much smaller than the sound travel time around the Sun, and therefore, local
effects are more important for these modes than for low-degree modes which
have much larger horizontal wavelengths and also longer life-times.

With the availability of data from the Michelson Doppler Imager (MDI) on
board the SOHO spacecraft, it is now possible to look at small areas
around a sunspot. In this work, we make a detailed comparative study of
the properties of high-degree oscillation modes in active and quiet
regions. We study how the different mode-characteristics
vary with the mean magnetic field of the
active regions.  
We use the ring diagram technique \citep{hill88} for our analysis.

The temporal variation of the mean frequencies, as well as other
characteristics of global modes of oscillations with solar cycle, has been
extensively studied \citep{elsworthetal90, libb-wood90, dziemboetal98,
bhatnagaretal99, howeetal99}. This variation is likely to be due
to the variation
in the global magnetic field over the solar cycle. The magnetic field in
sunspot regions is much stronger than the average field during maximum
activity, and hence, we expect a much larger effect if modes in active
regions are studied separately. This will also compensate for the larger
statistical errors of the ring diagram analysis technique as compared with global
mode analysis techniques, mainly due to
smaller spatial and temporal intervals covered by the ring diagram analysis.

Temporal variations in large scale flows using ring diagram analysis have
been studied extensively \citep{patron-etal98,basu-antia99,basu-antia00}.
\citet{haberetal01} have
studied both the temporal and spatial variation in flow velocities.  This
includes the variation in flow velocity with the magnetic field of active
regions. Although no clear systematic variation in flows with magnetic
field in active regions is found, \citet{haberetal01} claim that
horizontal flows are faster in active regions as compared to quiet
regions, and that flows converge around active regions.  
\citet{hindmanetal01} have tried to study variation in mode frequencies
with magnetic field. They find that mean frequencies of oscillations
increase with magnetic field in the region being considered. Furthermore, the
frequency shifts relative to the temporal and spatial average are found to
trace the active regions.

In this work, apart from frequencies and power variation we also include
other important characteristics of modes viz., width and asymmetry of peak
profiles. We also study variation in background power between pairs of
active and quiet regions.
The rest of the paper is organized as follows: we briefly discuss the analysis
technique and describe the data used in Section~\ref{sec:analysis},
present the results in Section~\ref{sec:results}, and discuss the results
and summarize our conclusions in Section~\ref{sec:conclu}.

\section{The Analysis} \label{sec:analysis}

Ring diagrams are three-dimensional (3D) power spectra of short-wavelength
modes in a small region of the Sun. High-degree (short wavelength) modes
can be approximated as plane waves over a small area of the Sun as long as
the horizontal wavelength of the modes is much smaller than the area
covered. Ring diagrams are obtained from a time series of Dopplergrams of
a specific area of the Sun tracked with the mean rotation velocity. The 3D
Fourier transform of this time series gives the power spectra. These power
spectra are often referred to as ring diagrams because of the
characteristic ring-like shape of the power in sections of constant
temporal frequency. A detailed description of the ring diagram technique
is given by \citet{patron-etal97} and \citet{basuetal99}.
Given the difficulty in
determining the properties of high-degree modes from global mode analyses
due to presence of leaks from adjoining $\ell,m$ values
(see e.g., Rhodes et al. 1998) and the fact that we are interested in
studying the mode properties in localized regions, ring diagrams provide
an alternative for such a study.

To extract the flow velocities and other mode parameters from the 3D power
spectra we fit a model for peak profile to these spectra.
The peak profiles are known to be asymmetric \citep{duvetal93,nigametal98}.
Although the exact form of
asymmetry is difficult to determine from the power spectra, the
model suggested by \citet{nigam-kosov98} is found to fit the
ring diagram spectra reasonably well \citep{basu-antia99}. 
Besides, use of asymmetric peak profile yields better agreement between
frequencies determined by fitting the velocity and intensity spectra
\citep{nigametal98,basuetal01}.
Thus we adopt the model for asymmetric peak profile as used
by \citet{basu-antia99} :
\be
P(k_x,k_y,\nu)= {\exp(A_0+(k-k_0)A_1+A_2({k_x\over k})^2+ A_3{k_xk_y\over
k^2})(S^2+(1+Sx)^2)\over x^2+1}+{e^{B_1}\over k^3}+ {e^{B_2}\over k^4},
\label{eq:model} \ee where \be x={\nu-ck^p-U_xk_x-U_yk_y\over
w_0+w_1(k-k_0)}, \label{eq:xx} \ee $k^2=k_x^2+k_y^2$, $k$ being the total
wave number, and the 13 parameters $A_0$, $A_1$, $A_2$, $A_3$, $c$, $p$,
$U_x$, $U_y$, $w_0$, $w_1$, $S$, $B_1$ and $B_2$ are determined by fitting
the spectra using a maximum likelihood approach \citep{andersonetal90}.
Here, $k_0$ is the central value of
$k$ in the fitting interval and $\exp(A_0)$ is the power in the ring, and
we refer to this quantity as peak-height or power in the rest of the text.
The coefficient $A_1$ accounts for the variation in power with $k$ in the
fitting interval, while $A_2$ and $A_3$ terms account for the variation of
power along the ring. The term $ck^p$ is the mean frequency, while
$U_xk_x$ and $U_yk_y$ represent the shift in frequency due to large scale
flows. The fitted values of $U_x$ and $U_y$ give the average flow velocity
over the region covered by the power spectrum and the depth range where
the corresponding mode is trapped. The mean half-width is given by $w_0$,
while $w_1$ takes care of the variation in half-width with $k$ in the
fitting interval. The terms involving $B_1,B_2$ define the background
power, which is assumed to be of the same form as that of
\citet{patron-etal97}. The asymmetry of the peak profiles is controlled by
the parameter $S$. The form of asymmetric profile is the same as that
prescribed by \citet{nigam-kosov98}. The quantities $U_x$ and $U_y$ can be
inverted to obtain the flow velocities in the east-west and north-south
directions respectively \citep{basuetal99}. The details of the fitting
procedure are explained in \citet{basuetal99} and \citet{basu-antia99}.
With asymmetric peak profile the calculated frequency ($\nu=ck^p$) does
not correspond to the point where power is maximum and the shift
with respect to the point of maximum power will depend on the line
width and the asymmetry parameter $S$.
The extent of this shift can be calculated from the assumed peak profile.

In this work we use the data obtained by the Michelson Doppler Imager
(MDI) on board the Solar and Heliospheric Observatory (SOHO) to determine
the mode characteristics and the flow velocities in the outer part of the
solar convection zone. We use the 3D spectra available in the MDI archives
for this work. These spectra were obtained from an area covering
$128\times128$ pixels, i.e., about $15^\circ\times 15^\circ$ in
heliographic longitude and latitude, giving a resolution of $0.03367$
Mm$^{-1}$ or 23.44 $R_\odot^{-1}$. The size of these regions is much
larger than that of a typical sunspot, hence the average magnetic field over the
region is smaller than the field inside the sunspots. Each region was
tracked for 1664 minutes.

In order to estimate the influence of magnetic field we compare the mode
characteristics in an active region with those in a quiet region at the
same latitude and in the same Carrington rotation. There are a number of
important reasons for this; the first being that the projection of the
spherical solar surface onto a flat one introduces some fore-shortening
that depends on the distance of the region from disk center, which can
introduce systematic errors in determining the mode characteristics. To
avoid this, we compare the results of an active region with that of a
quiet region at the same latitude. Each region in a pair is therefore
expected to have comparable effects of fore-shortening. In addition, to
minimize the effects of fore-shortening, we have only used spectra
obtained when the region under consideration was crossing the central
meridian. Another reason for comparing pairs of active and quiet regions
at the same latitude is that MDI images are known to have a certain amount
of distortion which changes with latitude and longitude. By comparing
regions at the same latitude and tracking them when they are at the
central meridian we minimize errors due to this distortion.

The time constraint on the regions compared is imposed because the optical
properties of the MDI instrument are known to change with time, resulting
in changes in the scale converting pixel size to physical length. Since
the scale length is needed to determine the mode properties in physical
units, we need to ensure that we only compare mode properties of regions
that have been analyzed within a small time interval.

For each of these regions we use a magnetic activity index (MAI) to denote
the mean strong field. This quantity was calculated by integrating the
unsigned field values
within the same regions and over the same intervals as those over which
the regions were tracked to calculate the power spectra, using available
96-minute magnetograms. The same temporal and spatial apodizations were
used. Only the strong field index was calculated. This was done by
setting all fields less than 50 Gauss to zero. One of the reasons of
calculating just the strong field index, rather than the total field
index is that the 96-minute magnetograms are a mix of 5-minute and 1-minute
averages. The averages do have slightly different zero levels and noise
levels, which can make a difference at low fields,  but is not expected 
to make much difference if the quantity of interest is only the 
strong field index.  In addition, to remove
effects of cosmic rays, outliers were removed. These were defined as
pixels with field values differing by a factor of more than six from the
average of their neighbors, if that average was more than 400 G.

We have analyzed eighteen pairs of regions. These regions cover a
reasonable range of latitudes within the active band and span a wide
interval of time (from about March 1996 to June 2000). The coordinates of 
the regions studied, as well as the
magnetic activity index of the regions, are tabulated in
Table.~\ref{tab:data}. We determine the mode parameters for each region by
fitting Eq.~(\ref{eq:model}), and we compare the parameters for each pair.
We believe that the difference in mode characteristics between an active
and quiet region at the same latitude and during the same Carrington
rotation arises from the influence of magnetic fields. Some differences
may also arise from temporal variations on short time-scales, but these
are expected to be small. This can also be tested by comparing the
characteristics in different quiet regions. In all cases we have found
that differences in mode characteristics between two quiet regions is much
less than the difference between a pair of active and quiet regions. Thus
we believe that most of the difference between an active and quiet region
characteristics is due to magnetic field.

\section{Results}
\label{sec:results}

We have studied four main properties of the oscillation modes, viz., frequency,
peak-height, line-width and the asymmetry parameter. We examine these
one by one.

\begin{figure}
\plotone{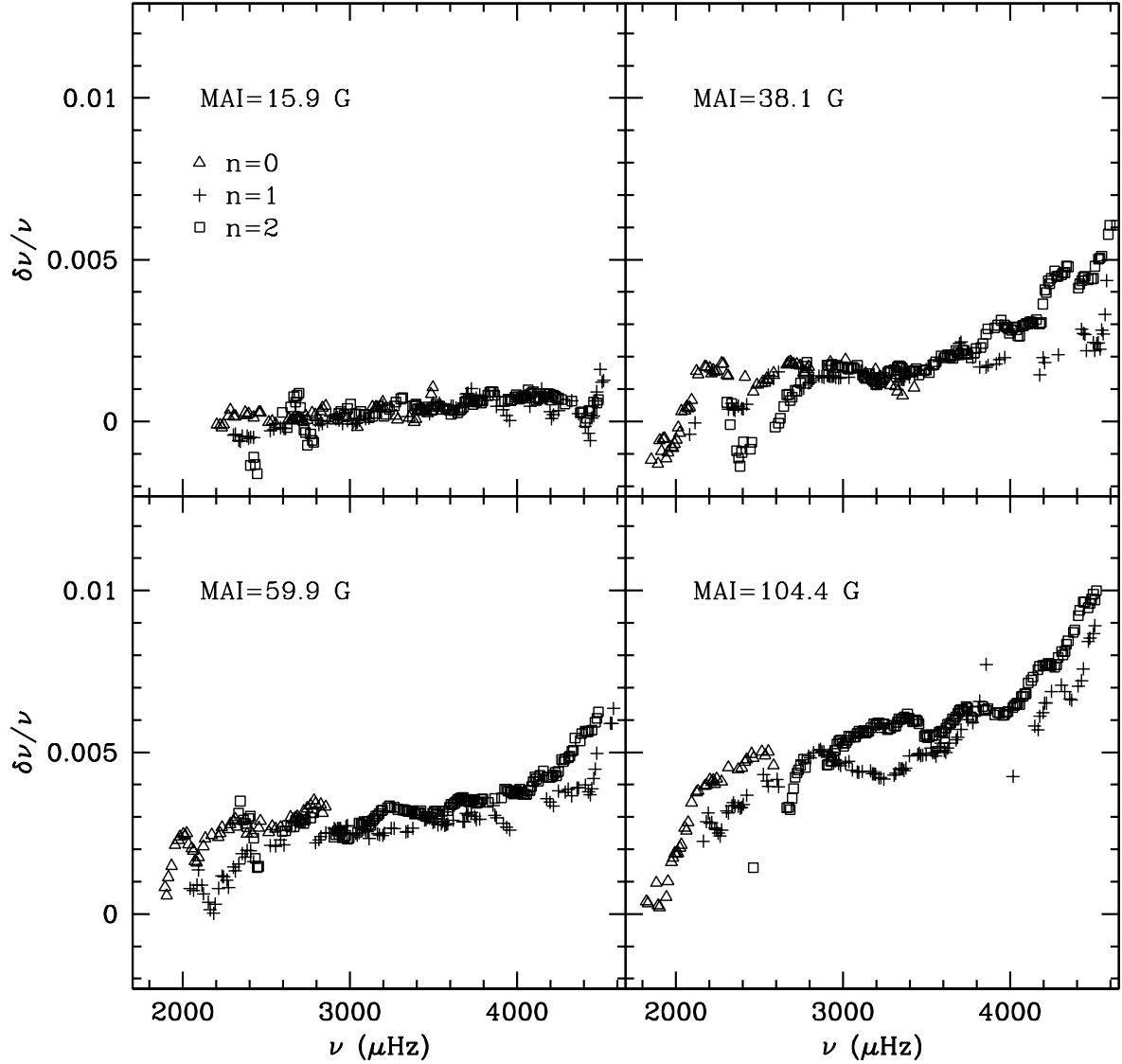}
\caption{The relative frequency differences as a function of frequency for four
different pairs of regions. The  error on the  points is about
0.0004 at 3.5 mHz and increases to 0.0007 above 4.5 mHz.
The MAI for the active region in each pair is indicated in the panels.}
\label{fig:dnu}
\end{figure}

Frequencies of both f  and p modes are higher in active regions than in
quiet regions.  This frequency shift is consistent with the expected
effect of magnetic field \citep{campbl-robs89,goldreichetal91}.
Fig.~\ref{fig:dnu} shows the
relative frequency differences, $\delta \nu/\nu = (\nu_{\rm
active}-\nu_{\rm quiet})/\nu_{\rm quiet}$, between a few pairs of active
and quiet regions. The mode frequencies are not only higher in active
regions, but they also increase with the magnetic activity index (MAI) of
the active region. The frequency differences increase with frequency and
become as high as 1\% for higher frequency modes, when the magnetic
activity index is high. For clarity the error-bars are not shown in
the figure, but the error is typically $0.0004$ at frequencies around
3.5 mHz and increases to $0.0007$ above 4.5 mHz.
Furthermore, the relative frequency shift is overall an
increasing function of frequency until about 5 mHz. We have
not shown the results beyond this frequency as the fits to the modes are
generally unreliable in these regions because of low power and large
widths. These high frequency modes are not expected to be trapped inside
the Sun.

\begin{figure} 
\plotone{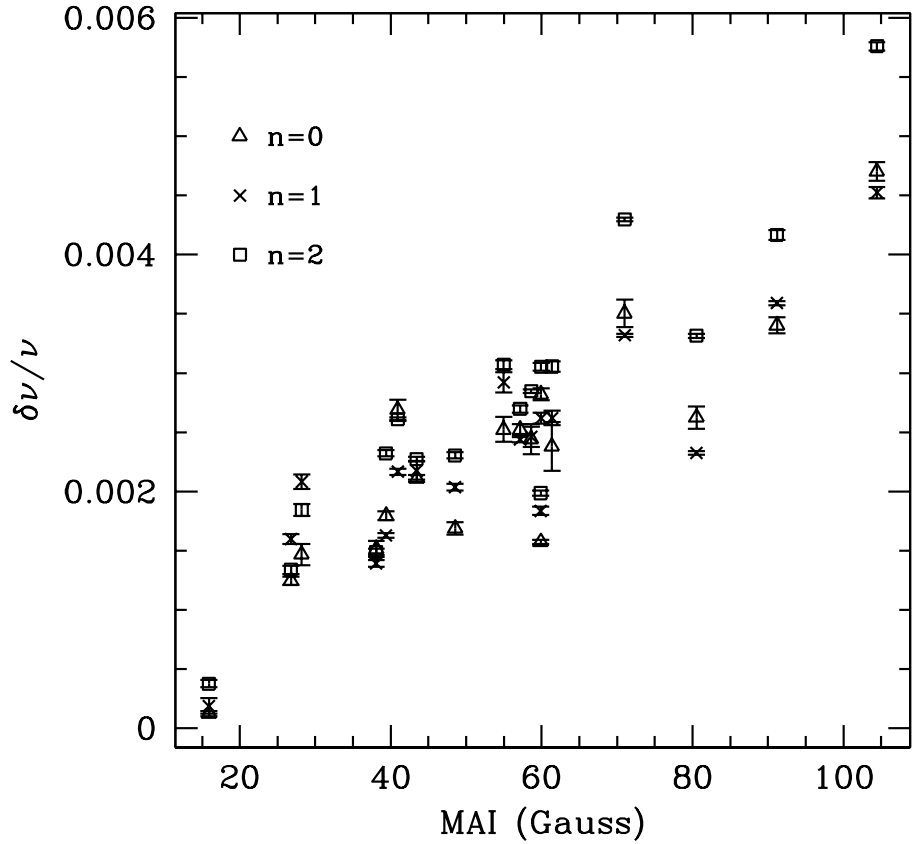} 
\caption{The average relative
frequency differences plotted as a function of the magnetic activity
index. Frequency differences of f modes have been averaged over modes in
the frequency range 2550 to 2750 $\mu$Hz, while those of p modes have been
averaged over modes in the range 3000 to 3500 $\mu$Hz. The averaging over
the relatively narrow frequency range is necessary because of the steep
frequency dependence of $\delta\nu/\nu$.} 
\label{fig:dnu_mai} 
\end{figure}

Fig.~\ref{fig:dnu_mai} shows the frequency-averaged relative frequency
differences between the active and quiet regions as a function of the MAI
of the active region. The relative frequency differences are averaged over
the frequency range of 2550 -- 2750 $\mu$Hz for f modes and 3000 -- 3500
$\mu$Hz for p modes.  The averaging over the relatively narrow frequency
range is necessary because of the frequency dependence of the frequency
differences. The correlation coefficients between $\delta \nu/\nu$ and the
MAI are 0.89, 0.88, 0.92, and 0.89, for $n=0$, 1, 2, and 3 modes
respectively. We note that the frequency differences are almost linear
with MAI, indicating that the strong field component of the magnetic field
of the regions under study has a major effect on the frequency
differences. The frequency shifts of f modes are comparable to those of
p modes, which further suggests a direct magnetic influence on the
oscillation modes, as any thermal perturbation would yield much lower
shifts for f modes as compared to that for p modes (e.g. Antia et al. 2001).

A striking feature of the frequency differences is that in all cases the
frequency differences appear to be roughly independent of $n$. It may be
noted that these frequency differences are not scaled by mode inertia,
and hence the frequency variations cannot be explained as arising entirely
from surface variations. A closer examination of Figs.~\ref{fig:dnu} and
\ref{fig:dnu_mai} shows that the frequency shifts slowly increase with the
radial order $n$, but this variation of $\delta \nu/\nu$ with $n$ cannot
be accounted for by the corresponding variation in mode inertia. Thus the
magnetic field must be penetrating below the outermost surface regions to
produce the observed variation. In order to get more information about
depth dependence of variation in Fig.~\ref{fig:dnurt} we show the same
frequency differences plotted as a function of lower turning point of the
modes. In this figure, the points for the different $n$ stand apart,
indicating that a major part of the frequency difference is a result of
perturbations close to the surface. The modes for each $n$ at high MAI
appear to show sharp variations, but this feature occurs at different
values of the lower turning point for modes of different radial order $n$.
This feature is less conspicuous in Fig.~\ref{fig:dnu} which shows the
same frequency differences as a function of frequencies. We discuss this
point and its implications in Section~\ref{sec:conclu}.

To show the variation in peak power we have plotted the ratio of the peak
power in the active and quiet regions as a function of frequency in
Fig.~\ref{fig:amp}.
The average error on each point is about 0.02 at about 3.5 mHz and
increases to about 0.07 above 4.5 mHz.
 We notice that the f and p modes in active regions in
general have lower power when compared with power in the quiet regions for
the same modes. The peak power in the active region can be as low as a
third of that in the quiet region for some modes.  The suppression
increases with increase in the MAI. Our results agree with early
observations which showed that power is reduced in active regions
\citep{tarbelletal88,hindman-brown98}.

The change in peak power is not monotonic with frequency. Maximum
suppression seems to occur in the frequency range of about 3 mHz to 3.5
mHz. This is similar to results of \citet{bogdanetal93} who found maximum
absorption around wavenumber $k\approx 0.8$ Mm$^{-1}$ or $\ell\approx
550$. This value of $\ell$ approximately agrees with the values where we
find maximum suppression of power for $n=1,2$ modes. It is not clear if
the mode absorption can be compared with suppression of power in the
sunspot region, since the absorption is measured for traveling waves,
while the modes we are studying are standing wave pattern. The suppression
of power decreases as the radial order $n$ of the modes increases. There
is a gradual shift of the position of maximum suppression towards higher
frequencies with increase in $n$. At high frequencies, the suppression
tends to become small and at low MAI it may change sign and become a power
enhancement instead. Similar behavior has been found by
\citet{hindman-brown98}.

Fig.~{\ref{fig:width}} shows the relative difference in half-width $w_0$
of the modes, between the active and quiet regions, plotted as a function
of frequency. 
On average the error on each point is about 0.02 at 3.5 mHz and
increases to about 0.04 above 4.5 mHz.
The half-width of a peak profile in the power spectrum is
related to the imaginary part of the frequency and hence is an indication
of mode-damping. The life-time of a mode is inversely proportional to its
half-width. The half-widths $w_0$ of the f  and p modes in the active
regions are generally larger than those in the quiet regions, at least for
$\nu > 2.5$ mHz. Thus it appears that most modes live longer in quiet
regions than they do in active regions, implying additional damping in
active regions. The change is generally larger for high magnetic field
regions. The maximum change in the width occurs in the frequency range 3
mHz to 3.5 mHz, quite similar to the behavior of maximum power
suppression. The change in the width seems to decrease with increasing
mode order, the f modes showing the largest change. A similar variation in
line-widths of global p modes has been found with solar cycle
\citep{kommetal00}. The variation in global mode widths is much less
($\approx 3\%$) than what we find, presumably because of smaller
variations in the average global magnetic field. Based on their
observations for global modes, \citet{kommetal00} predict an increase by
up to 38\% in widths and a reduction of up to 70\% in mode area in active
regions (the change in mode areas is shown in Fig.~\ref{fig:aw} and
discussed below). It is not clear what field strength they have assumed
for the active regions, but our results at high MAI are comparable to
their predictions. For example, for the highest MAI of $104.4$ G in our
sample, we find a maximum increase by about 50\% in widths
(Fig.~\ref{fig:width}) and a maximum reduction of about 55\% in mode area
(Fig.~\ref{fig:aw}), for $n$=1 mode.  The situation is not completely
clear for lower frequency modes. In any case it is difficult to determine
the width of low frequency modes reliably using ring diagram analysis,
since the actual width is smaller than the resolution of the power
spectra.

Given that the total power in the mode is the area under the peak in the
power spectrum, it is not instantly clear whether the increase in the line
widths in active regions compensates for the decrease in the peak height.
The area under a peak is a measure of excitation for the corresponding
mode. To check for this variation we have plotted the ratio of the product
of peak height and half-widths $A_0w_0$ as a proxy for the area under the
peak. The area should be a measure of acoustic power of the mode and hence
should contain information about mode excitation. This quantity is plotted
in Fig.~\ref{fig:aw}. The figures show that power in the modes is indeed
lower in high-activity regions. This is true even at low frequencies.
However, there seems to be some enhancement of power at higher frequencies
( $> 4$ mHz), particularly at low and intermediate magnetic activity
index.

The detection of the oscillation modes crucially depends on the
signal-to-noise ratio; hence it is important to compare the background
noise in the power-spectra, between the active and quiet regions.
The background in solar power-spectra is not completely noise, but has a
large component of so-called ``solar noise'', which is the background
produced by convective cells. The expression for the background in
Eq.~\ref{eq:model} has two terms with different $\ell$ dependences. We add
the two terms after taking the different $\ell$ dependence into account
(i.e., we calculate $\exp(B_1)\ell+\exp(B_2)$). The background is
predominantly a function of the degree $\ell$ of the mode, rather than the
frequency. We have plotted the ratio of background in the active and quiet
regions as a function of degree in Fig.~\ref{fig:bgl}. 
The average error on each point is about 0.07.
For most modes
the background in active region is less than that in quiet region.
It may be noted
that background should really be a function of $\nu$ and $\ell$, but since
we have determined it only in neighborhood of a peak we have labeled the
points by the corresponding value of $n$. At high frequencies it is
difficult to determine the background reliably, as the peak widths are
large. Hence the tails of the Lorentzian profiles extend quite far and
thus the `background' between two peaks is mostly contributed by the tails
of the peak profiles. That is why we have not shown the background at high
frequencies or high $n$. No significant variation in background power with
solar cycle has been found by \citet{kommetal00} in their analysis using
global modes.
The expected variation in the background due to small increase in average
magnetic field with solar cycle may be too small. Global mode analysis is
also restricted to $\ell<200$, where the variation is small even in our
results.

We can see from Fig.~\ref{fig:bgl} that the background in the active
regions is generally lower than that in the quiet regions. The reduction
may be expected because magnetic field is known to suppress convection,
which is the main source of background in power spectra. There is a steep
dependence of the background on the degree of the mode at intermediate
values of MAI; the situation at high MAI is not very clear. At
intermediate MAI, there appears to be a minimum in the background ratio
around $\ell=1000$, which corresponds to a horizontal wavelength of about
4 Mm. This length scale is somewhat larger than the granulation length
scale, but much smaller than the mesogranulation scale.

The asymmetry parameters for $n=0$ and $n=1$ modes for different regions
are plotted in Fig.~\ref{fig:asym}. 
The average error on each point is about 0.005 at 3.5 mHz.
The $n=2$ modes have not been plotted
for the sake of clarity; their behavior is very similar to that of the
$n=1$ modes. An asymmetry parameter of $S=0$ implies symmetric peaks. A
negative asymmetry implies that power in the low frequency half of the
peak is higher than that in the high frequency part. The larger the
absolute value of the asymmetry parameter, the more asymmetric is the
peak. We see that the asymmetry of the modes increases with increase in
the magnetic activity index of the region under consideration. This is
true for all the modes. The asymmetry of the line profile is generally an
indication of the depth at which the modes are excited \citep{kum-basu99}.
The deeper a mode is excited, the more symmetric is the peak. The increase
in the asymmetry of the profile can therefore be an indication of the fact
that the modes in the active regions are excited closer to the surface of
the Sun than the modes in the quiet regions.

As we did for the frequency shifts (Fig.~\ref{fig:dnu_mai}), we also study
the variation of the other mode parameters against the MAI of all the
active regions selected. The results are shown in Fig.~\ref{fig:dpar_mai}.
The frequency-averaged relative width differences are plotted in
Fig.~\ref{fig:dpar_mai}a. Although the results show a somewhat large
scatter, it is pretty clear that the line-widths increase with the
increase of MAI. The correlation coefficients between the relative
variation in width and the MAI turn out to be 0.49, 0.86, and 0.69 for
$n=0,1$, and $2$ respectively. The correlation is not strong, which is
reflected in the scatter. The scatter could indicate that the widths are
affected by the overall magnetic field and not merely the strong-field
component.

The frequency-averaged peak power ratios, as a function of the MAI, are
plotted in Fig.~\ref{fig:dpar_mai}b. We note the general trend of
increasing power suppression with MAI, however, there seems to be a
saturation effect at high activity index. The correlation coefficients
between the power ratio and MAI are $-0.56,-0.75,-0.58$ for $n=0,1,2$
respectively. Thus it is clear that variation in power is only weakly
correlated with MAI, probably because of saturation at high field.

Fig.~\ref{fig:dpar_mai}c shows how the frequency-averaged asymmetry
parameter varies as a function of the MAI. The increase in asymmetry seems
to be monotonic with MAI, though with some scatter. We note that
\citet{goode-strous96} (see also, \citet{rimmeleetal95}) have observed so
called ``seismic events'', which occur in the dark intergranular lanes and
are shown to excite the solar oscillations. They have shown that a local
magnetic field suppresses both the acoustic flux and the p-mode power.
These observations are of a quiet region, showing that even a weak
magnetic field significantly suppresses the acoustic energy flux due to
seismic events. Hence, our result that the mode asymmetry parameter
increases with the magnetic field in active regions implying excitation of
the modes closer to the surface might seem contradictory to the results of
\citet{goode-strous96} at first sight. But it should be noted that the
power in modes is reduced in the active regions, thus demonstrating that
excitation is reduced. This reduction in excitation probably shifts the
effective source of excitation upwards, as the reduction itself will also
have some depth dependence. It may also be possible that the magnetic
field induced activity at and above the surface layers, like the flaring
activities, may play a role in the excitation of oscillation modes in
active regions \citep{kosov-zhark98}, thus pushing the effective
excitation sources upwards.

The dependence of the degree-averaged background ratio on the MAI is shown
in Fig.~\ref{fig:dpar_mai}d. We note that the ratio seems to decrease with
MAI though there is a large scatter. The f-mode background however,
decrease reasonably monotonically with MAI. The correlation coefficient
between the background ratio and MAI is $-0.52$ for $n=0$ and $-0.40$ for
$n=1$. Thus it can be seen that the correlation between background and MAI
is rather weak as is apparent from the large scatter in the
Fig.~\ref{fig:dpar_mai}d.

Although flow velocity of solar plasma is not a mode characteristic, we
can determine the zonal and meridional flow components from the ring
diagram analysis. Unlike in the case of the other parameters, we are
unable to discern any systematic change in the flow velocities with
magnetic field and hence the results are not shown.  The flow velocities
are different in the quiet and active zones, but we cannot find a
systematic pattern in the changes with magnetic field.
Since we are effectively looking at averages over a region much larger
than the size of active region, the flows may also be averaged giving
more complex behavior. Haber et al.~(2001) have made more detailed
study of flows in active region using ring diagram analysis and they
claim a general increase in the flow velocities in active regions.
Our sample of regions studied is too limited to make detailed comments.

\section{Discussion and Conclusions}
\label{sec:conclu}

In this work we have performed plane-wave analyses of different active
regions of the Sun to study how mode characteristics of solar oscillations
change in these regions as compared with quiet regions at the same
latitude. We find that the characteristics of both f  and p modes in the
active regions are different from those in quiet regions. The change, in
most cases, varies with the MAI which is a measure of the average strong
magnetic field of the region under study.

\begin{figure}
\plotone{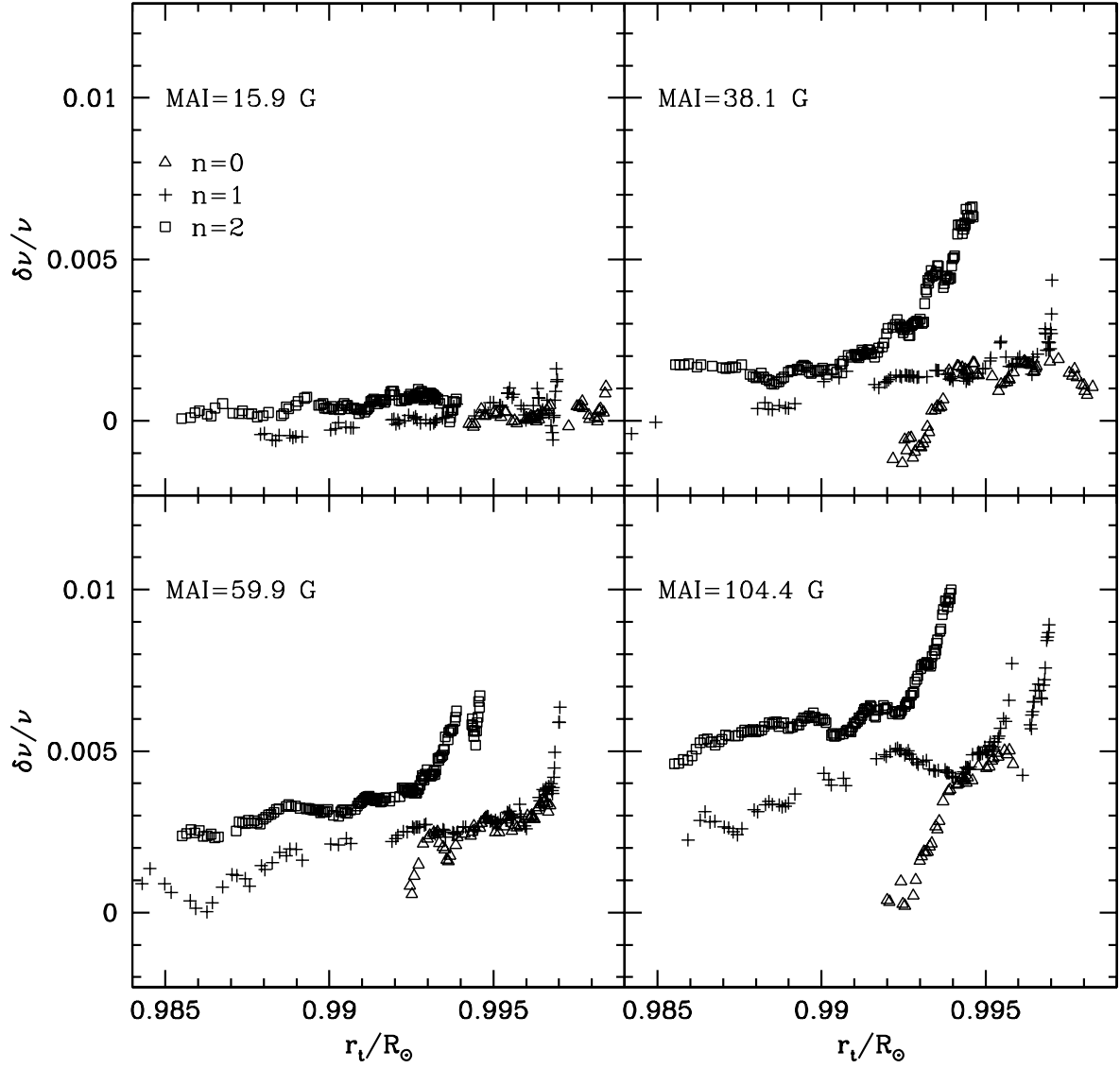}
\caption{The relative frequency differences as a function of lower
turning point of the modes, for four 
different pairs of regions.
The MAI for the active region in each pair is indicated in the panels.}
\label{fig:dnurt}
\end{figure}

\begin{figure}
\plotone{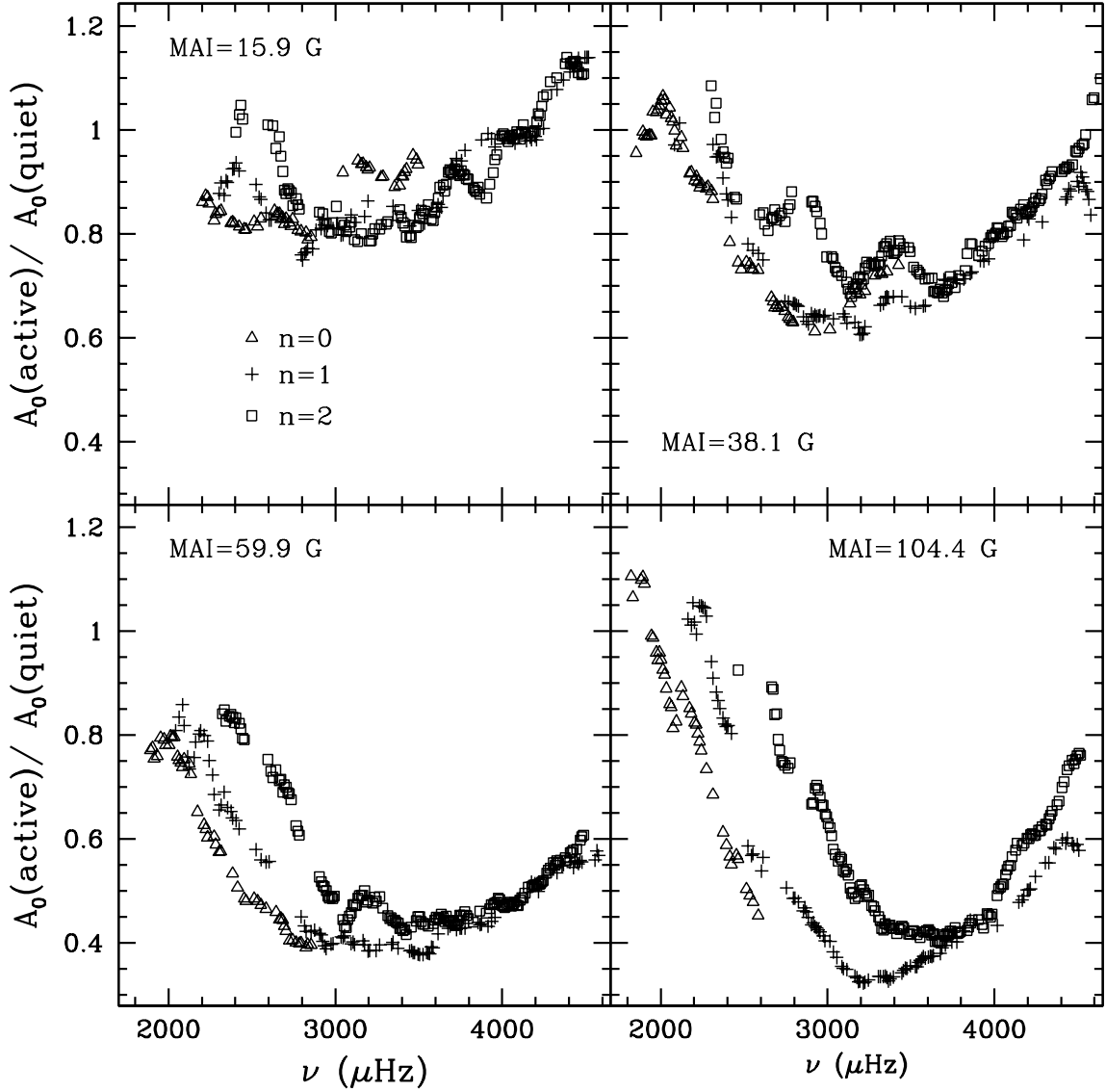}
\caption{The ratio of the peak power between pairs of active and
quiet regions.  The error on the points is about 0.02 at 3.5 mHz
and increases to 0.07 at 4.5 mHz.
The MAI of the active regions is indicated in the panels.}
\label{fig:amp} 
\end{figure}

\begin{figure}
\plotone{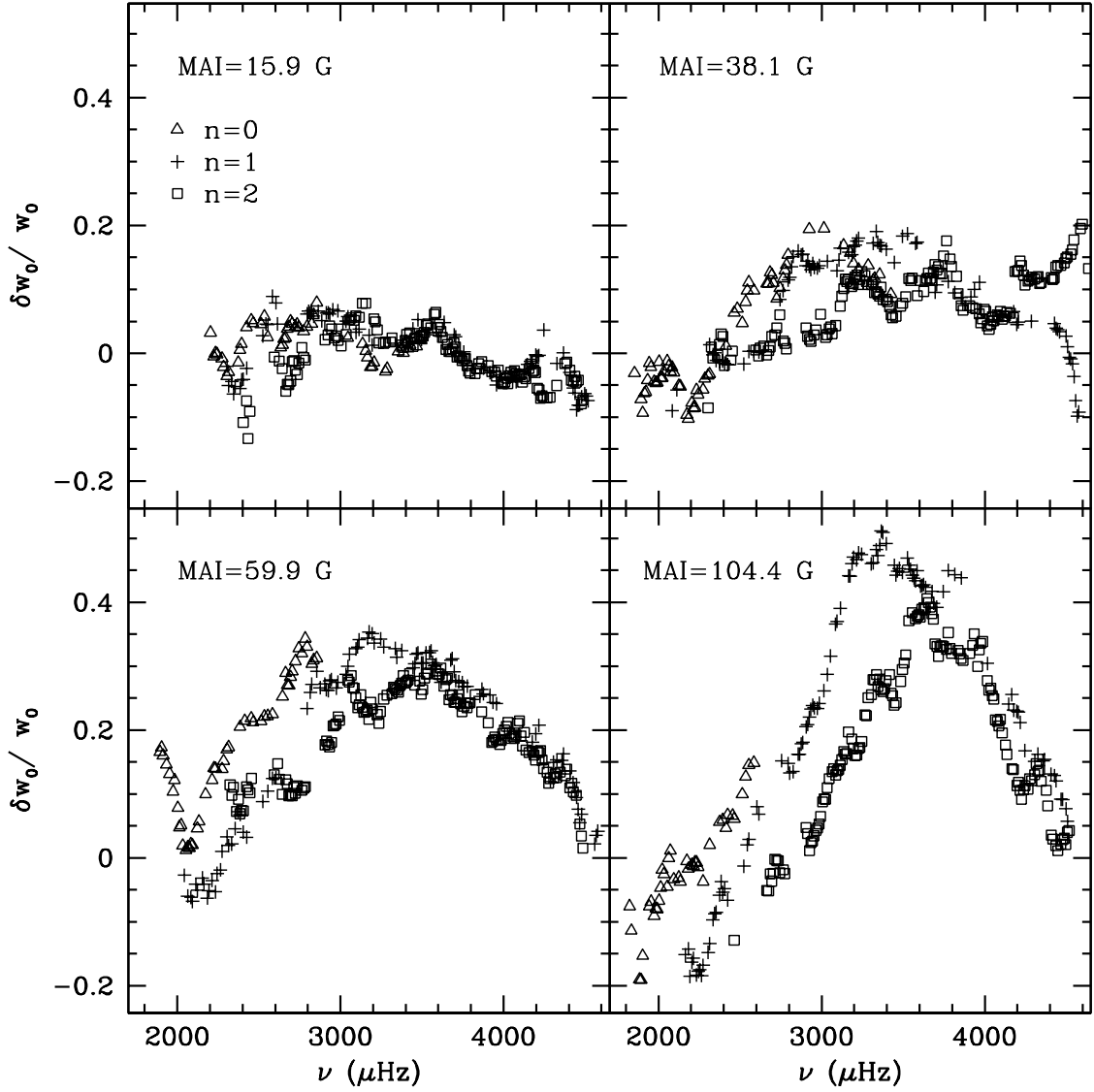}
\caption{The fractional differences in the mode line widths, 
$\delta w_{0}/w_{0}=(w_{0,\rm active}-w_{0,\rm quiet})/w_{0,\rm quiet}$,
as a function of frequency for four different pairs of regions.
The errors on the points increase from being about 0.02 at 3.5 mHz
to 0.04 above 4.5 mHz.
The MAI for the active region in each pair is indicated in the panels.}
\label{fig:width}
\end{figure}

\begin{figure}
\plotone{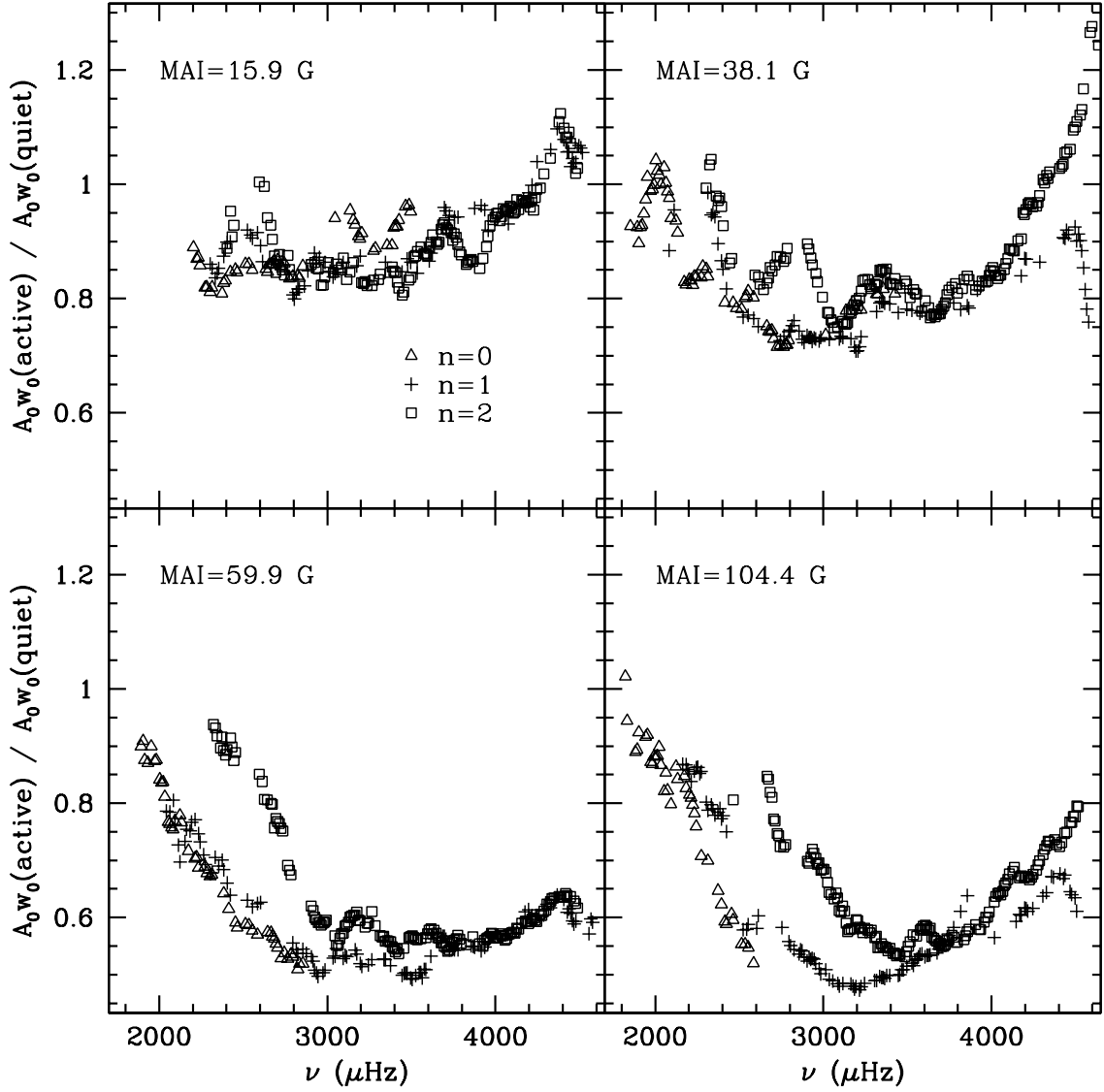}
\caption{The ratio of the product of the peak height and half-width
of the modes plotted as a function of frequency
for four different pairs of regions.
The MAI for the active region in each pair is indicated in the panels.}
\label{fig:aw}
\end{figure}

\begin{figure}
\plotone{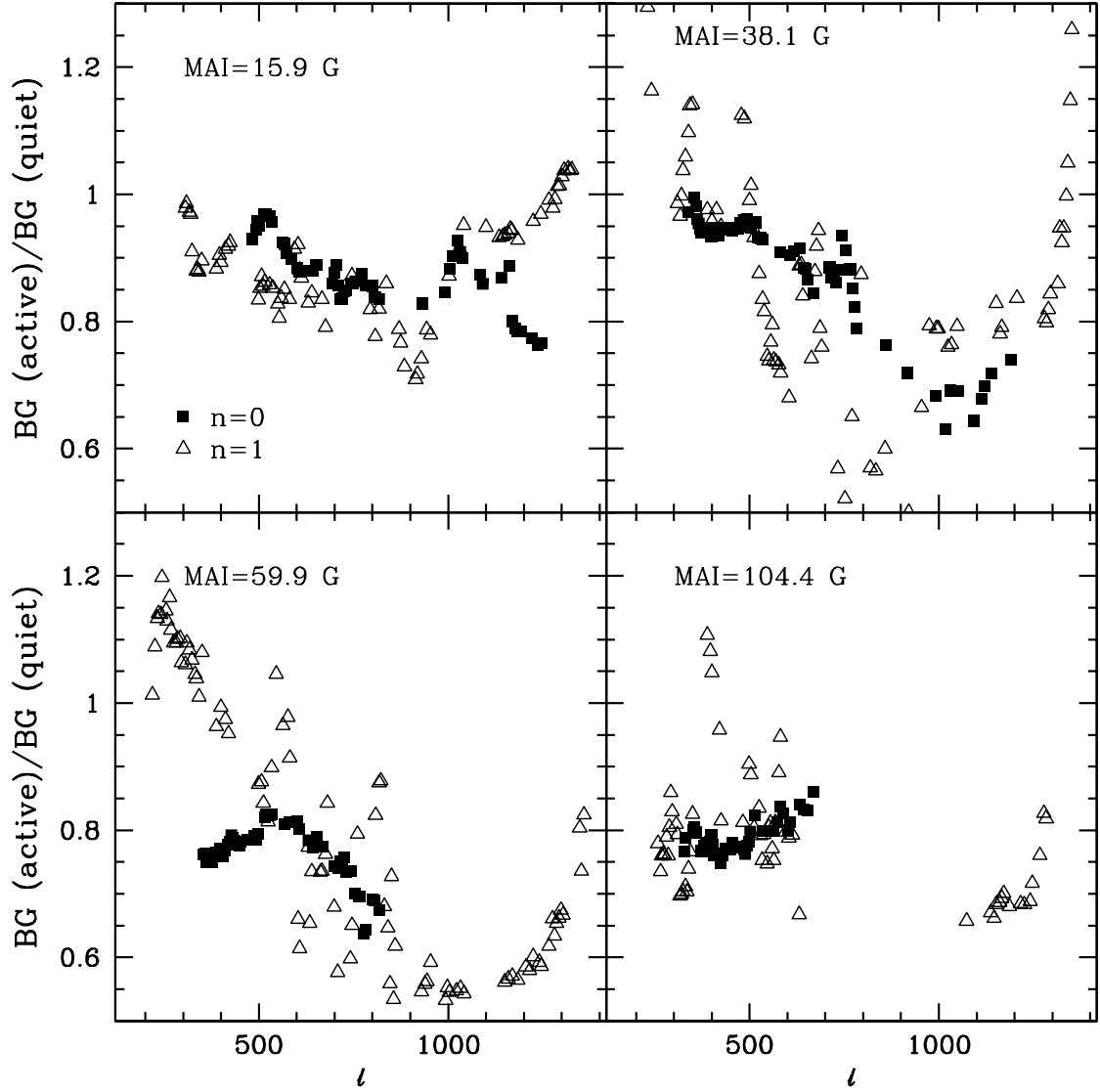}
\caption{The ratio of the background noise in the active and
quiet regions plotted as a function of degree
for four different pairs of regions. The errors on the points are
about 0.07.
The MAI for the active region in each pair is indicated in the panels.}
\label{fig:bgl}
\end{figure}

\begin{figure}
\plotone{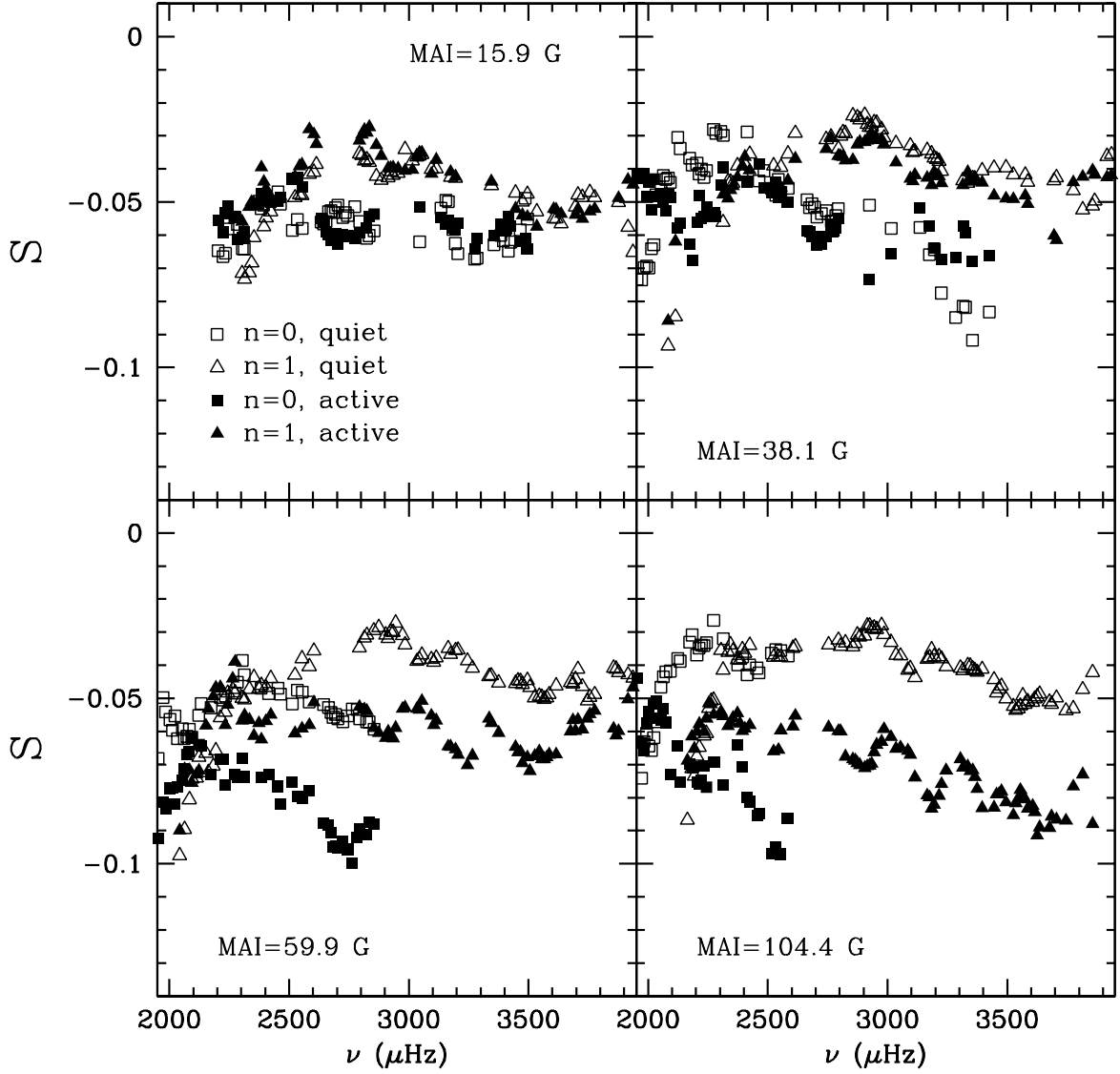}
\caption{The mode asymmetry parameter, $S$,
as a function of frequency for four different pairs of active and quiet
regions. The average error on each point is
about 0.005. The MAI of the active regions is indicated in the panels.} 
\label{fig:asym}
\end{figure}

\begin{figure}
\plotone{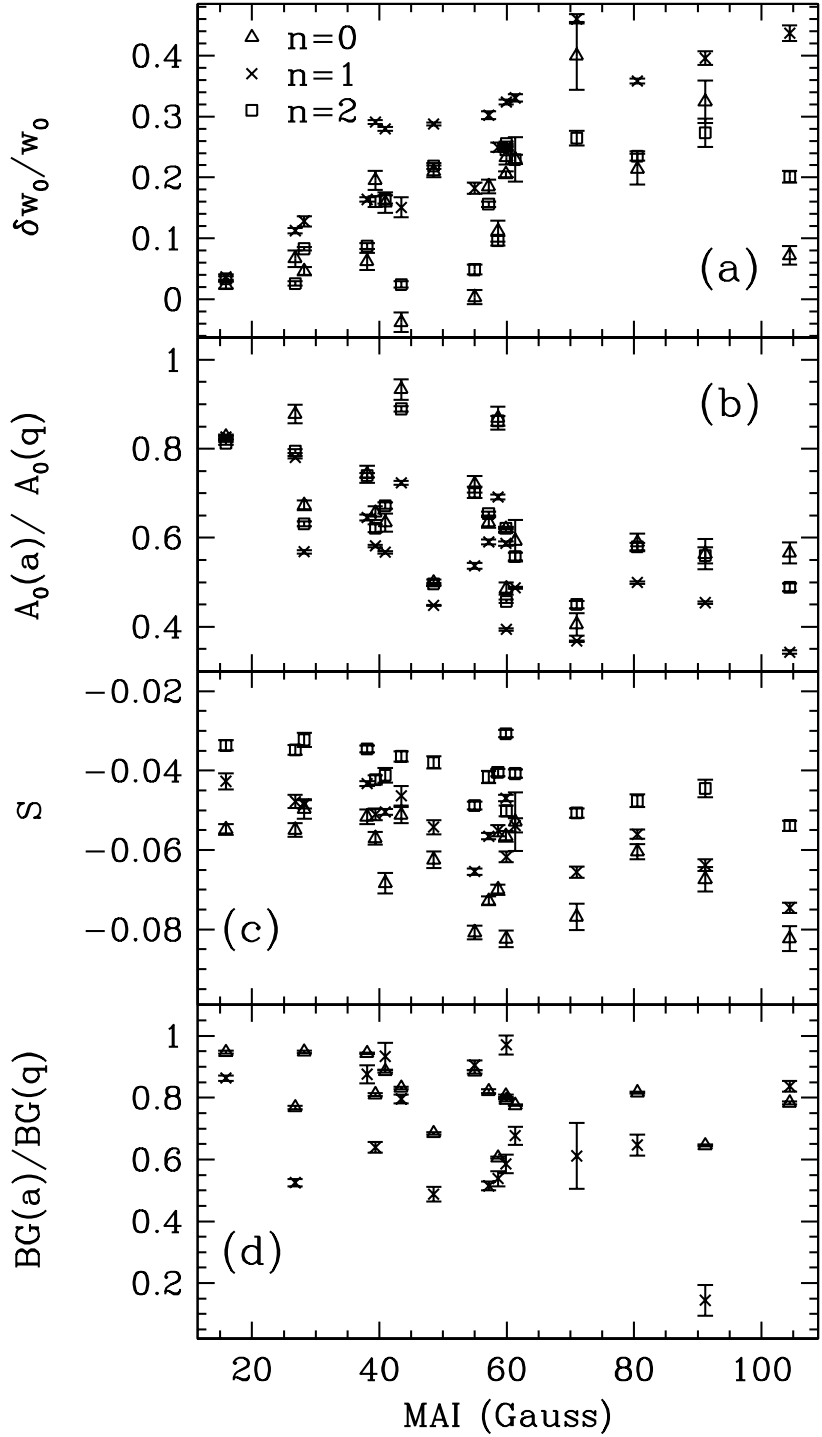}
\caption{The frequency-averaged mode parameters against the MAI of
the active regions; $a)$ fractional differences in the mode line widths,
$b)$ peak power ratios, $c)$ the asymmetry parameter $S$ and
$d)$ the ratio of the background noise in the active and quiet regions
averaged over the mode degree range $\ell=$400 to $\ell=600$.
The labels ``a'' and ``q'' in the axis labels imply active and
quiet regions respectively.}
\label{fig:dpar_mai}
\end{figure}

As with the solar cycle related variation of low and intermediate degree modes
\citep{libb-wood90}, the frequencies of high-degree modes increase with
increase in the mean magnetic field of the area under study.  The
frequency shifts of f  and p modes are comparable. This suggests that the
main cause of the frequency shifts is the magnetic field and not any kind
of change in solar structure. Changes in structure shift f mode
frequencies by much smaller amounts than they do p-mode frequencies.
Furthermore, it appears that the frequency shifts for modes of different
degrees cannot be explained as being due to differences in the mode
inertia of these modes. This gives us a handle on the penetration depth of
the magnetic fields. From Fig.~\ref{fig:dnurt} it can be seen that the
lower turning point of all modes considered in this study is below a depth
of 1.5 Mm from the solar surface. The magnetic fields must therefore
penetrate below this depth to explain the observed frequency variation,
which can not then be a purely surface effect. Considering that the
relative frequency differences are of the order of $10^{-2}$, the ratio of
magnetic to gas pressure in active regions should be of this order to
explain the observed frequency shifts. This gives a magnetic field of
about 100 G near the solar surface and correspondingly larger values in
the interior. The surface field is consistent with the actual mean value
inferred from the magnetograms in the region that we have considered.

It is, however, difficult to find a magnetic field configuration which
will yield observed frequency shifts characterized by a variation which
depend mostly on frequency alone without scaling by mode inertia. Wilson
depression in sunspots is another possible mechanism to explain the
frequency shifts and power reduction \citep{hindmanetal01}. Since the
effective depression can be expected to reduce with increasing depth, it
may explain the increase in frequency shift with frequency because the
higher frequency modes are reflected in higher layers. The depression can
also explain the comparable shifts in f- and p-mode frequencies since the
layers in which these modes are effectively trapped will also be
depressed. From Fig.~1 it can be seen that at highest field considered the
relative shift in f-mode frequency could be $0.005$, which will require
depression by about 2Mm. Depression by such a magnitude is probably not
expected. Thus it is quite possible that actual frequency shift is due to
a combination of direct effects of magnetic fields and of structural
variations like the Wilson depression. Any model to explain the frequency
shift should also be required to obtain f-mode shifts by amount comparable
to those for p modes.

The relative frequency shifts in active regions are found to have an 
overall increase with frequency up to a frequency of about 5 mHz. For modes
of intermediate degree, \citet{libb-wood90}, using data obtained at Big
Bear Solar Observatory (BBSO), found frequency shifts which have a maximum
around 4 mHz and then tend to reduce.  
It is difficult to fit individual modes in this frequency range due to
large widths and it becomes necessary to use ridge fitting techniques to
study these modes. The ridge fitting techniques have larger uncertainties
and it is not clear if the decrease in frequency shift seen in BBSO data
is real. Ring diagram analysis is also essentially a ridge fitting
technique but the spectra are not expected to be affected by leaks
arising from neighboring $\ell,m$ values.
We do not find any decrease in frequency shift with frequency.
Similar results have been found by \citet{hindmanetal01} also using ring
diagram analysis. Their results are consistent with ours.

The peak power of the modes in active regions is considerably smaller than
that in the quiet regions. This reduction in peak power seems to have
contributions from both acoustic absorption and reduced excitation in
active regions.  
Other possible mechanisms for the reduction of power in active
regions are
the modification of the  p-mode eigenfunctions
in magnetic region \citep{hindmanetal97} or  the fact that the height of 
formation of the spectral lines used to study oscillations is affected by 
magnetic field. 
A discussion of various mechanisms for the absorption of p-mode power 
can be found in \citet{spruit96}.

It is clear that there is a power enhancement in the high frequency modes
at low and intermediate field strengths, while the exact values of field
strength at which the transition to power reduction occurs seem to depend
on the radial order $n$ and it is not clear, though it is seen that higher
$n$ modes show larger power enhancement.  We note that such `halos' of
high-$\nu$ power enhancements for low and intermediate field strengths
have been reported in the literature \citep{hindman-brown98,thom-stanch00}
mainly at frequencies above the acoustic cut-off (5.0 mHz). This feature
has been suggested to be due to some mode of oscillation that is not
purely acoustic in nature, by \citet{thom-stanch00}, possibly due to
magnetic modes of oscillations in plage flux tubes that surround strongest
field spot regions \citep{hindman-brown98}. It has also been reported by
\citet{lindsey-braun99} using a different technique that sunspots are 
surrounded by acoustic ``halos''
of excess power at frequencies above 5 mHz, which they attribute to a
large scale eddy circulation induced by the convective energy blocked by
the sunspots.

The lower background power in the active regions seems to reinforce the
idea that high magnetic fields could hinder convection
\citep{bierman41,chandra61} which is the source of solar noise.  The
asymmetry in peak profile is believed to be an indicator of depth of
source of excitation, and an increase in asymmetry with magnetic field
implies a decrease in depth of source. The decrease in the excitation
depth with the increase in magnetic field implies that the turbulence
which excites the modes is pushed upwards towards the surface.  
\citet{goode-strous96} find suppression of acoustic flux and p-mode power
even in a weak magnetic field quiet region. We believe that magnetic field
induced activities in the surface and higher layers may play a role in the
excitation of oscillation modes in the active regions. Indeed, solar
flares have been observed \citep{kosov-zhark98} to excite seismic activity
in the active regions. The width of peak profiles also increase with
magnetic field, implying a shorter lifetime and hence higher damping due
to magnetic field. This is consistent with the results of
\citet{bogdanetal96}.

In conclusion we find that mode parameters change as a result of magnetic
fields in active regions. Judging by the frequency shifts, the influence
of the magnetic fields seems to penetrate to at least 1.5 Mm.
Unfortunately, these data are not yet precise enough to determine the
magnetic field by inversion. The power in the modes is smaller in active
regions than in quiet regions, while the modes are more asymmetric
indicating a shallower excitation depth.

\acknowledgments

We would like to thank Richard S. Bogart for providing the magnetic
activity indices.  This work utilizes data from the Solar Oscillations
Investigation/ Michelson Doppler Imager (SOI/MDI) on the Solar and
Heliospheric Observatory (SOHO). The MDI project is supported by NASA
grant NAG5-8878 to Stanford University. SOHO is a project of international
cooperation between ESA and NASA. This work was supported in part by NASA
Grant \# NAG5-10912 to SB.

\newpage

\begin{table}[ht]
\caption{Co-ordinates and Magnetic Activity Index of  Regions Analyzed 
\label{tab:data}}
\begin{tabular}{cccccc}
\tableline
CR \tablenotemark{\ast} & Latitude & CR Longitude & Magnetic Index &CR Longitude 
 &Magnetic Index\\
& & (active)& (active) & (quiet) & (quiet) \\
& (deg.) & (deg.) & (Gauss) & (deg.) & (Gauss)\\
\tableline
1933 & 22.5S & \ph 30 & \ph 91.19 & 180 & 0.28 \\
1934 &  22.5N &  285 & \ph 26.77 & 165 & 0.38 \\
1934 & 22.5S &  225 & \ph 48.53 & 315 & 0.99 \\
1934 & 22.5S & \ph 90 & \ph 39.38 & 165 & 0.44 \\
1947 &  22.5N &  270 & \ph 59.94 &\ph 30  & 3.61\\
1947 & 30.0S&  255 & \ph 28.21 &\ph 45  & 1.85\\
1948 &  22.5N &  285 & \ph 71.00 & 330 & 2.71 \\
1948 & 15.0S &  105 & \ph 40.94 &\ph 60  & 1.14 \\
1948 &  22.5N & \ph 30 & \ph 57.15 & 120 & 3.09\\
1949 &  22.5N &  300 & \ph 54.97 & 120 & 1.50 \\
1949 & 22.5S &  225 & \ph 43.44 &\ph 60  & 1.10  \\
1949 &  22.5N &  210 & \ph 58.62 &\ph 60  & 5.22 \\
1949 & 15.0S &  120 & \ph 38.09 &\ph 60  & 0.80\\
1949 &  30.0N & \ph 90 & \ph 15.91 & 120 & 2.11\\
1963 & 15.0S &  180 & \ph 59.88 & 105 & 1.89\\
1963 &  22.5N & \ph 75 &  104.45 & 225 & 1.80 \\
1964 &  22.5N &  285 & \ph 80.52 & 105 & $\cdots$ \\
1964 & 15.0S &  120 & \ph 61.37 &\ph 90 & 2.98 \\
\end{tabular}
\tablenotetext{\ast}{Carrington Rotation}
\end{table}

\newpage

\end{document}